\documentclass[preprint,amsmath,amssymb,aip,groupedaddress]{revtex4-1}
\usepackage{graphicx}
\usepackage{dcolumn}
\usepackage{bm}
\begin{document}

\title{Enhancement of superconducting transition temperature in electrochemically etched FeSe/LaAlO$_3$ films}

\newcommand{\Tc}{$T_{\mathrm c}$ }
\newcommand{\Ts}{$T_{\mathrm s}$ }
\newcommand{\IG}{$I_{\mathrm g}$ }
\newcommand{\VG}{$V_{\mathrm g}$ }
\newcommand{\FeSeS}{FeSe$_{1-x}$S$_x$}
\newcommand{\FeSeTe}{FeSe$_{1-y}$Te$_y$}
\newcommand{\Tczero}{$T_{\mathrm c}^{\mathrm {zero}}$ }
\newcommand{\Tconset}{$T_{\mathrm c}^{\mathrm {onset}}$ }
\author{N. Shikama}
\email[]{shikama@g.ecc.u-tokyo.ac.jp}
\author{Y. Sakishita}
\author{F. Nabeshima}
\author{Y. Katayama}
\author{K. Ueno}
\author{A. Maeda}
\affiliation{Department of Basic Science, University of Tokyo\\
Meguro, Tokyo 153-8902, Japan}
\date{\today}
\begin{abstract}
In this study, we investigated the gate voltage dependence of \Tc in electrochemically etched FeSe films with an electric-double layer transistor structure. The \Tczero value of the etched FeSe films with a lower gate voltage (\VG = 2.5 and 3.3 V) reaches 46 K, which is the highest value among almost all reported values from the resistivity measurements except for the data by Ge {\it et al}. This enhanced \Tc remains unchanged even after the discharge process, unlike the results for electrostatic doping without an etching process. Our results suggest that the origin of the increase in \Tc is not electrostatic doping but rather the electrochemical reaction at the surface of an etched films.
\end{abstract}

\maketitle
\newpage
\section{Introduction}
Iron-based superconductors have attracted significant attention owing to their high superconducting transition temperature ($T_{\mathrm c}$). FeSe \cite{FeSe_Hsu} (\Tc = 9 K), the material with the simplest crystal structure among all iron-based superconductors, has shown an increase in \Tc of up to 30-50 K by the doping of electrons through an intercalation of ions when using a solid-state reaction method,\cite{KFe2Se2_Guo} an ammonothermal method,\cite{MetalIC_Ying} an ion-exchange method, \cite{LiFe_Dong,Niu_LiFe_ARPES} and an electrochemical reaction method. \cite{Krzton-Maziopa_IC,S.Hosono_IC,CTA_Shi,TBA_Shi} The electrostatic doping of electrons into FeSe using the electric-double layer transistor (EDLT) structure of the ionic-liquid gates also enhances superconductivity. \cite{EDLT_Lei,EDLT_Hanzawa} In FeSe-EDLT, the Hall coefficient abruptly changes from positive to negative at a specific gate voltage $V_{\mathrm g}$, indicating the occurrence of a Lifshitz transition, leading to an increase in \Tc of up to 30 K. With a further increase in $V_{\mathrm g}$, the \Tc of electron-doped FeSe increases up to 48 K. \cite{EDLT_Lei} 

Shiogai {\it et al}. achieved control of the thickness and doping level control using a single FeSe-EDLT device through electrochemical etching, and investigated the thickness dependence of \Tc on electron-doped FeSe films. \cite{EDLT_Shiogai}
They reported a critical thickness of 6-10 nm, depending on the substrate material, and found that for films thinner than the critical value, the \Tc values were approximately 40 K.
However, in our previous study, \cite{EDLT_Kouno} we determined that there is no such critical thickness, and even for films of thicker than 50 nm, \Tc increases up to 40 K once the  dead layer of the film surface (the thickness of which is approximately 5-10 nm) is removed. We also suggested the existence of an electrochemical reaction layer at the surface of etched FeSe films which is responsible for the high \Tc of 40 K, because a change in resistance is irreversible even after the discharge and recharge processes. If the origin of the high \Tc in the  etched FeSe is the existence of the electrochemical reaction layer, the relation between the carrier density and the gate voltage of the etched FeSe will differ from that of the electrostatic-doped FeSe because the concentration of the doped electrons will depend on how the electrochemical reaction layer is formed. Thus, detailed features of such phenomena, for instance, how \Tc depends on $V_{\mathrm g}$, might be complicated when compared with simple electrostatic doping.

In this study, we investigated the gate voltage dependence of the superconducting properties of etched FeSe/LaAlO$_3$ (LAO) films with an EDLT structure to clarify the mechanism of the \Tc enhancement in etched FeSe-EDLT. During this process, we obtained a further increase in $T_{\mathrm c}$ for lower \VG values than typical values for etching (\VG = 5.0 V), and the etched FeSe films with \VG = 2.5 V showed \Tconset = 50 K and \Tczero = 46 K with good reproducibility, the latter of which is higher than the \Tczero obtained from any other resistivity measurements of FeSe, except for the results by Ge {\it et al}. \cite{ML_Ge} This high \Tc persists even when \VG is removed. This, together with the characteristic dependence of \Tc on $V_{\mathrm g}$, strongly suggests that the increase in \Tc of the etched FeSe films originates not from electrostatic doping, but from the electrochemical reaction. \cite{EDLT_Kouno}
\section{Method}
All FeSe films applied were grown on LaAlO$_3$(LAO) substrates using a pulsed laser deposition (PLD) method with a KrF laser, and the EDLT structure was fabricated on the grown FeSe films in the air. Details of the film growth and fabrication of the EDLT structure are described elsewhere. \cite{ImaiJJAP,ImaiAPEX,FeSe_strain_Kawai,EDLT_Kouno} Here, N,N-dethyl-N-methyl-N-(2-methoxyethyl)ammonium bis(trifluoromethanesulfonyl)imide (DEME-TFSI) \cite{EDLT_Hanzawa,EDLT_Shiogai,EDLT_Lei,EDLT_Kouno} was used as the gate dielectric. The thickness of the pristine films, as measured using a Dektak stylus profiler, was 30-60 nm. X-ray diffraction measurements of the pristine films showed no impurity phase. Resistivity measurements and the etching processes of the FeSe films with an EDLT structure were conducted in a He atmosphere using a Quantum Design Physical Property Measurement System (PPMS) at temperatures of 3 to 330 K.
\section{Results and Discussion}
First, we confirmed that the pristine FeSe films showed a superconducting transition at 3-5 K, which was lower than the \Tc of the bulk sample owing to the tensile strain. \cite{FeSe_strain_Kawai} Then, the target gate voltage $V_{\mathrm g}$ was applied at $T$ = 220 K, just above the glass transition temperature of DEME-TFSI. The $V_{\mathrm g}$ values in this study were 2.5, 3.3, 5.0 and 5.5 V. As mentioned above, based on the {\it ex-situ} experiments, to obtain an enhanced $T_{\mathrm c}$, we removed the dead layer of the surface of the films through electrochemical etching. The temperature was raised with the applied gate voltage maintained so that electrochemical etching takes place. The etching temperature depends on $V_{\mathrm g}$, and was 330 K for \VG = 2.5 and 3.3 V, and 240 K for \VG = 5.0 and 5.5 V. The FeSe films could not be etched under a \VG of less than 2.5 V because the FeSe films started to degrade at temperatures higher than 360 K, where etching is induced for such low \VG values.
%
%
\begin{figure}
 \centering
 \includegraphics[width=8.5cm,clip]
      {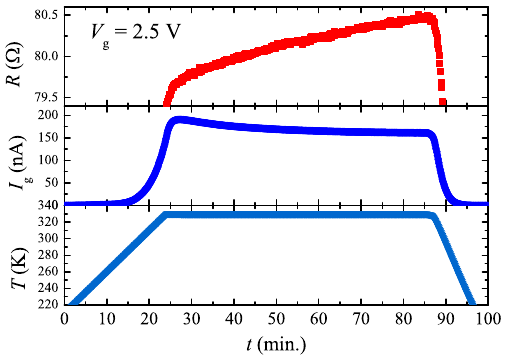}
\caption{Temporal variation of the resistance, $R$, and the gate current, $I_{\mathrm g}$, during a typical etching process at a gate voltage of \VG = 2.5 V.}
\label{Etprocess}
\end{figure}

Figure \ref{Etprocess} shows the typical etching process at \VG = 2.5 V. The gate current, $I_{\mathrm g}$, increased gradually as the temperature, $T$, increased, which indicates that the etching started.  We observed a slight increase in resistance when $T$ was kept constant. This indicates that the thickness of the films decreased by etching. After etching the films by a few nanometers, we stopped each etching process by decreasing the temperature. (We estimated the thickness of the etched FeSe films from the Faraday charge,  $Q_\mathrm{F}$ = $\int$\IG d$t$. \cite{EDLT_Shiogai,EDLT_Miyakawa,EDLT_Kouno})
\begin{figure}
 \centering
 \includegraphics[width=7.5cm,clip]
      {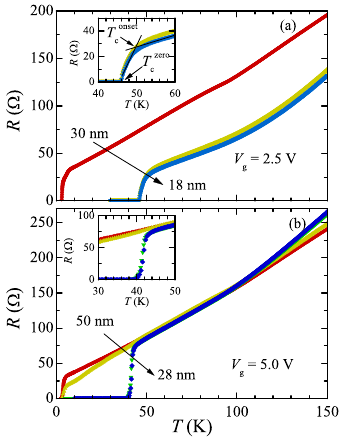}
\caption{Temperature dependence of the resistance of etched FeSe films for (a) \VG = 2.5 V and (b) 5.0 V. The insets show magnified views of the plots near the superconducting transition temperature after the etching process.}
\label{RT_et}
\end{figure}

Figure \ref{RT_et} shows the changes in temperature dependence of the resistance when the etching processes for \VG = 2.5 and 5.0 V were repeated. The value of \Tc increased up to 40-50 K for \VG = 2.5 and 5.0 V after repeating the etching process, and remained constant after the further etching process. This behavior is essentially the same as that shown in the previous results. \cite{EDLT_Shiogai,EDLT_Kouno,EDLT_Miyakawa} Note that the steeper drop of the resistance was observed as compared with those for other FeSe-EDLT \cite{EDLT_Hanzawa,EDLT_Lei,EDLT_Shiogai,EDLT_Miyakawa,EDLT_Kouno} and that \Tczero for \VG = 2.5 V was 46 K. This is remarkable, because \Tczero = 46 K observed for \VG = 2.5 V is the highest \Tczero among all previously reported values obtained based on the resistivity measurements of FeSe. This includes the resistivity measurement of monolayer FeSe films, except for the result of Ge {\it et al.} (\Tc of $\sim$ 100 K), \cite{ML_Ge} the reliability and reproducibility of which have been the subject of debate. \cite{Bozovic_ML_rev}
\begin{figure}
 \centering
 \includegraphics[width=8.5cm,clip]
      {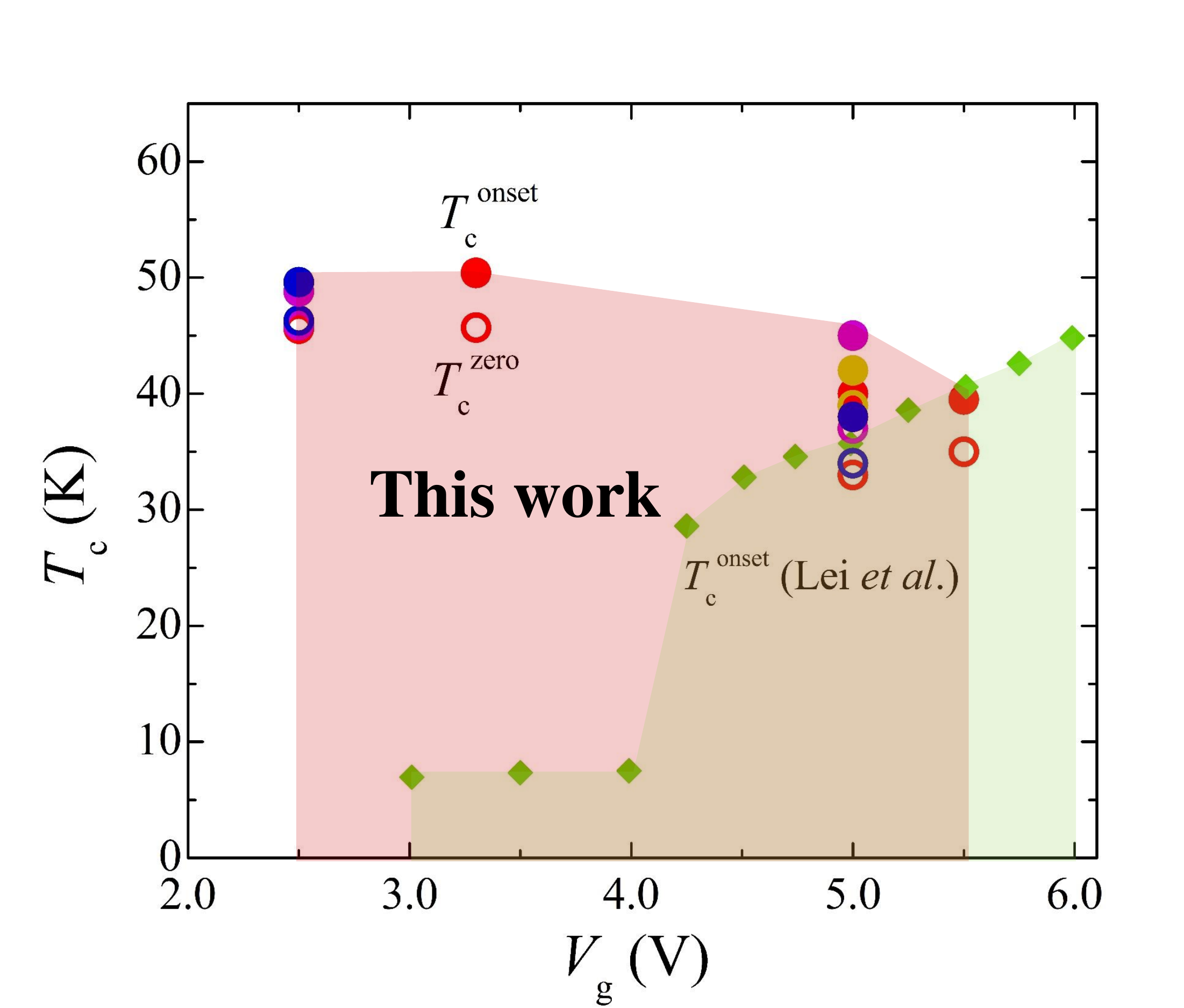}
\caption{Gate voltage dependence of \Tc for FeSe films on LAO. The closed and empty circles represent \Tconset and $T_{\mathrm c}^{\mathrm {zero}}$, respectively. The data of the several samples for \VG = 2.5 and 5.0 V are plotted in different colors. The closed diamond represents the data on electrostatically doped FeSe single crystal from Lei {\it et al}. \cite{EDLT_Lei}}
\label{PD}
\end{figure}

Figure \ref{PD} shows the \VG dependence of \Tc on the etched films. We also plotted the \Tc data of electrostatically doped FeSe by Lei {\it et al}. \cite{EDLT_Lei} for comparison. We repeated the experiments for several samples. The \Tc data obtained are also plotted in Fig. \ref{PD}. The highest \Tc of 50 K was observed for \VG = 2.5 and 3.3 V with good reproducibility. The enhancement of \Tc with a lower \VG is distinctly different from the results of the field-effect study for bulk FeSe, \cite{EDLT_Lei} in which \Tc increases drastically up to 30 K at \VG = 4.25 V and gradually increases to 48 K for a further increase in $V_{\mathrm g}$. This strongly suggests that the mechanism of \Tc enhancement of etched FeSe is different from that of electrostatically doped FeSe.
\begin{figure}
 \centering
 \includegraphics[width=8.0cm,clip]
      {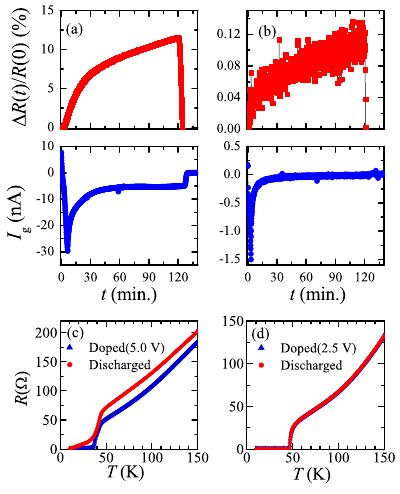}
\caption{(a),(b) Temporal variation of $R$ and \IG in the discharge process at 220 K for \VG = 5.0 V and 2.5 V, respectively. Here, $t = 0$ min represents the time when the discharge process started by decreasing \VG to 0 V. In addition, $\Delta R(t) = R(t)-R(0)$ represents the change in resistance from $t = 0$ min. (c),(d) Temperature dependence of the resistance before (blue) and after (red) the discharge process for FeSe films etched at \VG = 5.0 V and 2.5 V, respectively. The data for \VG = 2.5 V are almost overlapped.}
\label{discharge}
\end{figure}

Figure \ref{discharge}(a) shows the change in the resistance and \IG before and after the discharge process for \VG = 5.0 V. During the discharge process, we decrease \VG to 0 V at 220 K to avoid unexpected additional etching at higher temperatures, and kept the temperatures at 220 K for two hours with \VG = 0 V. After \VG was returned to 0 V, the discharge current was observed, which gradually decreased over time. We also observed a gradual increase in $R$, which might be due to the release of the induced electrons and/or the decomposition or removal of the surface conducting layer. \cite{EDLT_Kouno} Qualitatively, the same behavior was observed for \VG = 2.5 V (Fig. \ref{discharge}(b)), but the absolute value of \IG and the increase in the resistance are much lower than those for \VG = 5.0 V. Two hours after \VG was removed, the temperature dependence of the resistance of the FeSe films was measured again. Although \Tczero decreased down to 10 K, a high \Tconset was still observed for \VG = 5.0 V after the discharge process (Fig. \ref{discharge}(c)).
Furthermore, for \VG = 2.5 V, both \Tconset and $T_{\mathrm c}^{\mathrm {zero}}$ and the temperature dependence of the resistance remained unchanged (Fig. \ref{discharge}(d)).
These results strongly suggest that the electrons are introduced not by electrostatic doping but by the electrochemical reaction between the ionic liquid and the FeSe because \Tc of the electrostatically doped FeSe decreases to less than 10 K when the gate voltage is decreased back to 0 V. \cite{EDLT_Lei}

The data in Fig. \ref{discharge} shows remarkable differences in and during the discharge process between \VG = 5.0 V and \VG = 2.5 V. Let us discuss the origin of the difference in terms of the amount of the discharge charge. We can consider two different contributions for the discharge. One is the discharge of electrostatically doped charge, and the other is the charge originated from the possible decomposition process, as will be discussed below.  
The amount of the charge which flowed during the two-hour-discharge process per unit area was approximately 110 $\mu$C/mm$^2$ for \VG = 5.0 V, which was much larger than the expected electrostatic charge for an EDLT with DEME$^+$ (0.1-1 $\mu$C/mm$^2$ at 5.0 V). This suggests that the dominant contribution of the discharge for \VG = 5.0 V is the decomposition. \cite{EDLT_Kouno} An electrochemical reaction layer with thickness of tens of nanometers, which is responsible for the high $T_{\mathrm c}$, was decomposed during the discharge process, resulting in a decrease in $T_{\mathrm c}$. On the other hand, the discharge charge amount for \VG = 2.5 V was 1.0 $\mu$C/mm$^2$, which is much smaller than that for \VG = 5.0 V, and comparable to the expected electrostatic charge. This strongly suggests that the electrochemical reaction layer formed at \VG = 2.5 V hardly decomposed through the discharge process at 220 K. This is consistent with the observation that the \Tc value and the $R$-$T$ curves remained unchanged after the discharge process. This also means that the electrostatic doping (1.0 $\mu$C/mm$^2$ at most) did not affect $T_{\mathrm c}$ for the etched FeSe for \VG of 2.5 V. This suggests that the electrochemical reaction dopes much more carriers than the electrostatic doping.
\begin{figure}
 \centering
 \includegraphics[width=8.5cm,clip]
      {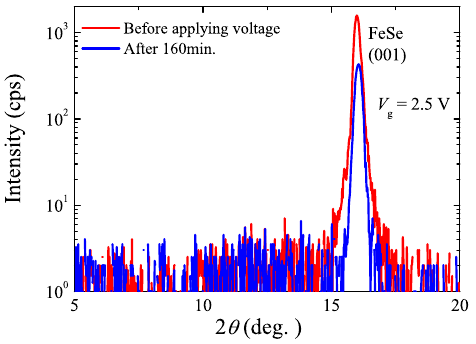}
\caption{Temporal variation of the X-ray diffraction patterns of an FeSe film with the EDLT structure during the electrochemical etching at \VG = 2.5 V at room temperature in air for 5\(^\circ\)$<$ $2\theta$ $<$ 20\(^\circ\). XRD scans were performed before and after the gate voltage was applied, which takes 30 min to finish an XRD scan.}
\label{OoP}
\end{figure}

If the origin of the electrochemical reaction is an intercalation, the DEME$^+$ cation should be intercalated into FeSe layers. \cite{EDLT_Kouno} To clarify if the intercalation is induced in bulk, we conducted X-ray measurements. Figure \ref{OoP} shows the X-ray diffraction (XRD) patterns with \VG = 2.5 V. A reflection peak should be observed at 2$\theta$ = 5\(^\circ\)-10\(^\circ\) if an increase in the $c$-axis length by the intercalation of the DEME$^+$ cation takes place. \cite{KFe2Se2_Guo,Niu_LiFe_ARPES,Krzton-Maziopa_IC,S.Hosono_IC,MetalIC_Ying,LiFe_Dong,CTA_Shi,TBA_Shi} Although the peak intensity of the (001) reflection of FeSe was suppressed by the etching of the films, no additional peak was observed during the etching process. There are two possibilities for the absence of additional peaks. One possibility is that the thickness of the layers where the DEME$^+$ cation was intercalated is too thin for any additional peaks to be observed. The other possibility is that the DEME$^+$ cation is adsorbed only at the surface of the FeSe, which cannot be detected using XRD. Although further studies are needed to determine which possibility is true, the FeSe layers with an enhanced \Tc are considered extremely thin.

Together with the results of the discharge experiments, the XRD results suggest that the electrochemical reaction layer for \VG = 2.5 V was very thin and stable, while that for \VG = 5.0 V was much thicker and unstable. This may suggest that different electrochemical reactions take place between \VG = 2.5 V and 5.0 V; for example, only the adsorption of DEME$^+$ takes place for \VG = 2.5 V, while the intercalation of DEME$^+$ in addition to the DEME$^+$ adsorption is induced for \VG = 5.0 V.  The characteristic \VG dependence of \Tc for the etched films might be understood as follows. The DEME$^+$-adsorption layer shows higher \Tc than that for the DEME$^+$-intercalated layer, but \Tc of the adsorption layer for \VG = 5.0 V is suppressed because the surface adsorption layer is deteriorated by the intercalation of the large DEME$^+$ cation to the layers underneath.

High \Tc values have recently been reported in organic ion-intercalated FeSe compared with other intercalated compounds. Cetyltrimethyl ammonium ions (CTA$^+$) \cite{CTA_Shi} and tetrabutyl ammonium ions (TBA$^+$) \cite{TBA_Shi} intercalated with FeSe show a \Tconset of 45 and 50 K, respectively, which are relatively higher than those of metal-intercalated FeSe, \cite{KFe2Se2_Guo,Niu_LiFe_ARPES,Krzton-Maziopa_IC,S.Hosono_IC,MetalIC_Ying,LiFe_Dong} but comparable with our \Tc discussed above. This suggests that the high \Tc of 50 K is a characteristic phenomenon for an electrochemical reaction without metal. We speculate that there are fewer formations of impurity phases and disorders because of the inertness of the organic molecule ions, as suggested by Shi {\it et al}. \cite{CTA_Shi,TBA_Shi} In etched FeSe-EDLT systems, another enhancement of \Tc was reported using multivalent ionic liquids, \cite{EDLT_Miyakawa} as compared with monovalent DEME-TFSI. This may suggest that the optimization of \VG for EDLT using multivalent ionic liquids \cite{EDLT_Miyakawa} can further increase the \Tc of Fe-chalcogenides above 50 K.
\section{conclusion}
In this study, we investigated the gate voltage dependence of the superconducting transition temperature of etched FeSe films. Enhancement of \Tc at lower gate voltages (\Tconset = 50 K, \Tczero = 46 K) was observed with good reproducibility. In addition, \Tczero = 46 K is the highest reported value from the resistivity measurements of FeSe, except for the result by Ge {\it et al}. \cite{ML_Ge} The XRD and discharge experiments suggest that the enhanced \Tc in the etched FeSe for lower \VG values originates from the DEME$^+$ adsorption to the surface of FeSe. Our results indicate that finding the best condition for the electrochemical reaction of FeSe can be the key to a further enhancement of $T_{\mathrm c}$.
\begin{acknowledgments}
This research was supported by JSPS KAKENHI Grant Numbers 18H04212 and 19K14651 and by the Precise Measurement Technology Promotion Foundation (PMTP-F).
\end{acknowledgments}
\section*{Data Availability Statements}
The data that support the findings of this study are available from the corresponding author upon reasonable request.

\end{document}